\documentclass[review]{elsarticle}

\usepackage{lineno,hyperref}
\usepackage{amsmath,bm}
\usepackage{lscape,color}
\usepackage{multirow}
\usepackage{xcolor}
\usepackage{hhline}

\definecolor{byzantine}{rgb}{0.74, 0.2, 0.64}
\modulolinenumbers[5]

\journal{Elsevier}
\begin{document}
	
	\begin{frontmatter}
		\title{Bayesian estimation of the autocovariance of a model error in time series}
		\author{Yoon Bae Jun}
		\ead{junpeea@gmail.com}
		\ead[url]{https://github.com/junpeea}
		\address{Iowa State University, United States}
		
		\author{Chae Young Lim}		
		\ead{twinwood@snu.ac.kr}
		\address{Seoul National University, Korea}

		\author{Kun Ho Kim\corref{mycorrespondingauthor}}
		\cortext[mycorrespondingauthor]{Corresponding author}
		\ead{kunhokim8@gmail.com}
		\address{Concordia University, Canada}
		
		\begin{abstract}
		Autocovariance of the error term in a time series model plays a key role in the estimation and inference for the model that it belongs to. Typically, some arbitrary parametric structure is assumed upon the error to simplify the estimation, which inevitably introduces potential model-misspecification. We thus conduct nonparametric estimation of it. To avoid the difficult bandwidth selection issue under the traditional nonparametric truncation approach, this paper conducts the Bayesian estimation of its spectral density in a frequency domain. To this end, we consider two cases: fixed error variance and time-varying one. Each approach is taken to estimate the spectral density of the autocovariance and the model parameters. The methodology is applied to exchange rate forecasting and proves to compete favorably against some benchmark models, including the random walk without drift.
		
		\end{abstract}

		
		
	\end{frontmatter}
	
	\linenumbers
	 
\section{Introduction}
\label{sec:introduction}

   Heteroskedasticity has played a key role in economic time series analysis. It is well-known that a number of important macroeconomic or financial return processes exhibit the pattern of heteroskedasticity (Engle, 1982) \cite{E82}. It has been one of the central issues in the literature how to model such processes to capture their underlying dynamics successfully. Classical time-varying volatility models, such as $ARCH$ and $GARCH$, that accommodate conditional heteroskedasticity are typically stationary with some restrictions on the parameter values. The $ARCH$ and $GARCH$ processes are unconditionally uncorrelated while they are still dependent. For the uncorrelated error, the corresponding spectral density is constant and estimating unconditional variability of classical $ARCH$- or $GARCH$-type models may not be of interest. To that end, one can incorporate a heteroskedastic time-varying volatility within the model where the error terms are not only dependent but also unconditionally correlated. For example, see Kim and Kim (2016) \cite{K16}. In this study, we illustrate how such a framework can be useful in practice.
		
	 Specifically, we consider the following framework:
\begin{eqnarray}
\label{eqn:model}	
\bm{y} = \bm{X}\bm{\beta} + \bm {\epsilon},
\end{eqnarray}
   where $\bm{y} = (y_1,\cdots,y_T)'$ be a variable of interest, $\bm{X} = (X_1;\cdots;X_p)$ be a covariate matrix where $X_i$ is a ($T\times 1$) vector of the $i$-th predictor, $\bm{\beta}$ is a ($p\times 1$) vector of unknown regression parameters and $\bm \epsilon=(\epsilon_1,\cdots, \epsilon_T)'$ is an error vector.  Here we assume $\epsilon_t  = \sigma_{\epsilon,t} \, e_t$, where $e_t$ is a weakly stationary and stochastically continuous Gaussian process with $E(e_t)=0$ and $Var(e_t)=1$ and $\sigma_{\epsilon,t}^2$ is a time-varying volatility of $\epsilon_t$. Here $\sigma_{\epsilon,t}^2$ is also a marginal variance function of the error $\epsilon_t$. We denote $\gamma(\cdot)$ by the autocorrelation function for $e_t$. Inference for the autocorrelation will be carried out in a frequency domain by utilizing the corresponding spectral density, $\lambda(\cdot)$.  In such a setting, a log transformation of the time-varying volatility $\sigma_{\epsilon,t}^2$ can be represented by $B$-spline functions. Then, a hierarchical Bayesian framework can be adopted such that some appropriate prior distributions can yield posterior distributions of the spectral density and of the time-varying volatility as well as those for the regression parameters.

   As shown in Dey {\it et al.} (2018) \cite{dey2018bayesian}, adoption of a Gaussian process prior for the log transformation of a spectral density not only helps capture the time dependence structure in the error process, but also enhances the precision of forecasting through a better forecast of the model error. The current study employs the prior choice for the spectral density suggested by Dey {\it et al.} (2018) \cite{dey2018bayesian}, but also extends their work by allowing the volatility in \eqref{eqn:model} to change over time. It is well-known that a model with heteroskedasticity captures the underlying dynamics of an economic/financial time series variable more efficiently than a naive constant-volatility model. By adopting a flexible modeling for the volatility of such a process, one can expect potentially more precise estimation and forecasting for the variable of interest.
	
   Estimation of the autocorrelation function for the error process via the spectral density and of the related model parameters is conducted under a Bayesian framework. In particular, the hierarchical Bayesian framework is known to be useful for macro-finance forecasting (Kim, 2011 \cite{K11}; Dey {\it et al.}, 2018 \cite{dey2018bayesian}). Moving into the frequency domain from the time domain, we use an approximate likelihood for the parameters including the spectral density. 
	
	 Unlike the traditional time-domain-based one, the $frequency$-domain-based approach adopted in this study allows us to avoid the difficult issue of how to choose a nonparametric bandwidth for the truncation of long-run variance. In addition, the approach ensures the Bayesian estimate of the covariance matrix to be {\it positive}-{\it definite}, which is a non-trivial advantage in a data-rich environment. Furthermore, the proposed Bayesian estimation does not assume a specific parametric form for any time-dependent structure, by allowing one to incorporate relevant information into the covariance matrix estimation through the prior distribution of the spectral density, which helps reduce the likelihood of model mis-specification. 
	
   The methodology proposed in this study is applied to the issue of {\it exchange rate forecasting}. One of the crying needs in the modern macro-finance literature is to develop a methodology or a model that can forecast foreign exchange rates with decent accuracy (Rossi, 2013 \cite{Ros13}). Over the years, many approaches with various degrees of success have been proposed in the literature. Meese and Rogoff (1983) \cite{MR83} compare out-of-sample forecasting accuracy of various exchange rate models and find out that a random walk model outperforms the rest at one- to twelve-month horizons for the dollar/pound, dollar/mark and dollar/yen. Engel and Hamilton (1990) \cite{EH90} develop a long-swing model that explains the uni-directional move in the dollar value for long periods of time. They claim that their model generates better forecasts than the random walk process. 
	
	 Mark (1995) \cite{M95} shows that fundamentals, such as money stock and output, are useful in predicting exchange rates and that the regression with bias-adjusted coefficients can outperform a random walk process in forecasting the spot exchange rate. Kilian (1999) \cite{K99} reports no evidence of higher predictability for exchange rates at longer forecast horizons and concludes that the linear model framework underlying many empirical studies are likely to be mis-specified, suggestive of a non-linear data-generating process. To that end, Kilian and Taylor (2003) \cite{KT03} propose a non-linear, exponential smooth transition autoregressive model to approximate the time series behavior of real exchange rate. Molodtsova and Papell (2009) \cite{MP09} find strong evidence of exchange rate predictability with the Taylor rule fundamentals at the one-month horizon for 12 OECD countries over the post-Bretton Woods period. Ince {\it et al.} (2016) \cite{IMP16} further show that the Taylor-rule-fundamental model outperforms the Taylor-rule-differential model (Engel {\it et al.}, 2008) \cite{EMW08} and other traditional models in out-of-sample exchange rate forecasting. 	

   On the Bayesian side, Wright (2008) \cite{W08} suggests pooling forecasts from several different exchange-rate models, called the Bayesian model averaging, for the problem of pseudo-out-of-sample exchange rate predictions. He shows that the forecasts generated by the model averaging methodology are very close to those based on the random walk process. Through the flexible nonparametric modeling approach proposed in this paper, the frequency-domain-based Bayesian estimation of the error auto-covariance and the time-varying volatility would yield further progress in forecasting exchange rate variables. 
	
   The paper is organized as follows: Section 2 introduces some background relevant to our proposal. Specifically, we present particulars of the proposed nonparametric modeling of the error covariance and of the time-varying-volatility approach. The estimation procedure is described as well. A simulation study supporting our approach and a real data example involving foreign exchange rate data are provided in Section 3. The paper ends with its conclusion and discussion on potential extensions in Section 4. Details of the proposed methodology are relegated to an appendix.

\section{Proposed Method}
\label{sec:methods}
	
Recall that we consider the time series regression setting given in \eqref{eqn:model} with $\epsilon_t=\sigma_{\epsilon, t} e_t$:
\begin{eqnarray}
\label{eqn:model2}	
\bm{y} = \bm{X}\bm{\beta} + \bm {\sigma}_t\odot \bm e,
\end{eqnarray}	
where we let $\bm \sigma_{\epsilon}=(\sigma_{\epsilon, 1},\cdots, \sigma_{\epsilon, T})'$ and $\bm e=(e_1,\cdots, e_T)'$ to explicitly decompose the error into a {\it time}-{\it varying} volatility and a normalized stationary error process. Here $\odot$ represents an element-wise multiplication. We further assume that $e_t$ follows the stationary Gaussian process with zero mean, unit variance and autocorrelation function $\gamma(h)$. 

Meese and Rogoff (1983) \cite{MR83} and Rossi (2013) \cite{Ros13} mention {\it instabilities} as one of the potential reasons why traditional economic models with time-invariant parameters have not been successful in out-of-sample forecasting of exchange rates. Our framework in (\ref{eqn:model2}) partially addresses the issue by introducing the time-varying volatility, which could potentially yield more precise forecasts of exchange rates than the traditional ones.

Given the assumption of weakly stationarity for $e_t$ in our model \eqref{eqn:model2}, Wold's theorem (Priestley, 1981 \cite{priestley1981spectral}) states that the function $\gamma(\cdot)$ is an autocorrelation function, equivalent to the spectral measure $\Lambda(w)$ such that:  
\begin{equation}
\label{sec:2.1.1.1}
\gamma(h) = \int_{-\pi}^{\pi} \exp(iw h)\Lambda(dw)
\end{equation}
where $i=\sqrt{-1}$ and $\Lambda(w)$ is a distribution function on $[-\pi,\pi)$. When $\Lambda(w)$ is differentiable everywhere, the spectral density function $\lambda(w)=\frac{d\Lambda(w)}{dw}$ exists and \eqref{sec:2.1.1.1} becomes 
\begin{equation}
\label{eq:spectraldensity}
\gamma(h) = \int_{-\pi}^{\pi} \exp(iw h)\Lambda(dw)=\int_{-\pi}^{\pi} \exp(iw h)\lambda(w)dw.
\end{equation}
 The spectral density function $\lambda(w)$ is recovered by the Fourier transformation of $\gamma(\cdot)$. That is,
$\lambda(w) = \frac{1}{2\pi}\sum_{h=-\infty}^{\infty} \gamma(h) \exp(-i h w) $ for $w \in [-\pi,\pi)$.
Following the relation in \eqref{eq:spectraldensity}, modeling the spectral density $\lambda(w)$  is equivalent to modeling the autocorrelation function $\gamma(h)$. 

It is well-known that the truncated version of the Fourier transformation of a sample autocorrelation yields an estimate of the spectral density, which is called the {\it periodogram}. The periodogram is defined by:
\begin{equation}
\label{sec:2.1.2.4}
\mathcal{I}_{n}(w) = \frac{1}{2n\pi}\left|\sum^{n}_{t=1}e_{t}\text{exp}(-itw)\right|^2,  \nonumber
\end{equation}
\noindent where $w \in [-\pi,\pi)$.  
From the Gaussianity assumption on $\bm e$,   $\mathcal{I}_n(w)$ is asymptotically exponentially distributed with mean $\lambda(w)$ and $\mathcal{I}_n$ are asymptotically independent at Fourier frequencies, $w_j = 2\pi j/n$, for $j=- \lfloor \frac{n}{2} \rfloor ,\cdots,\lfloor \frac{n-1}{2} \rfloor$; $\lfloor x \rfloor$ being the greatest integer less than or equal to $x$. Since $\lambda(-w)=\lambda(w)$ as well as $\mathcal{I}_n(-w)=\mathcal{I}_n(w)$, we will only consider $j=1, \cdots, \lfloor \frac{n-1}{2} \rfloor (=m)$. 
Carter and Kohn(1997) \cite{carter1997semiparametric} utilize that  the logarithm of a standard exponential random variable can be approximated by a mixture of five Gaussian random variables to have the following $approximated$ relationship:
\begin{equation}
\label{sec:2.1.2.6}
\log \mathcal{I}_{n}(w) = \log \lambda(w) + \xi(w)
\end{equation}
with $\xi$ having a distribution $\pi(\xi)$ such that 
\begin{equation}
\label{sec:2.1.2.7}
\pi(\xi) \propto \sum_{l=1}^{5}p_l\phi_{v_l}(\xi-k_l),
\end{equation}
\noindent where $\phi_v(\xi - k)$ is normal density with mean $k$ and variance $v^2$. Here $(p_l, k_l, v_l)$ for $l=1,\cdots, 5$ are known and are provided by Carter and Kohn (1997) \cite{carter1997semiparametric}. 

We use \eqref{sec:2.1.2.6} at Fourier frequencies to obtain an approximate likelihood for the spectral density in Bayesian inference.  
That is, the autocorrelation function $\gamma (h)$ for $\bm e$ is then implicitly considered in 
$
\bm{\varphi} = \bm{\theta}+ \bm{\xi}
$
 in the frequency domain, where  $\bm{\varphi}=(\log(\mathcal{I}_n(w_1)),\cdots, \log(\mathcal{I}_n(w_m)))'$  is a vector of the logarithm of the periodogram and $\bm{\theta}=(\log(\lambda(w_1)), \cdots, \log(\lambda(w_m)))'$  is a vector of the logarithm of the spectral density at Fourier frequencies. Here $\bm{\xi}$ is a vector of the five component mixture Gaussian random variables at Fourier frequencies from \eqref{sec:2.1.2.6}.  
Moreover, $\log(\sigma_{\epsilon, t}^2) \equiv \eta_{\epsilon, t}$ is represented by cubic $B$-spline basis functions such that $\eta_{\epsilon, t}=\sum_{b=1}^d \delta_b \phi_b(t)$.

In this model setting, the parameters to be estimated are $\bm \beta$, $\bm \delta=(\delta_1, \cdots, \delta_d)'$, $\bm \theta$ as well as  latent label information, denoted by $\bm \psi$, for $\bm \xi$. For the Bayesian inference, we consider prior specifications of the parameters as follows: For $\bm \beta$, we consider  $\bm\beta \sim N(\mu_{\beta} \bm 1, \sigma_{\beta}^2\bm I)$, which is a multivariate Gaussian distribution with mean vector $\mu_{\beta}\bm 1$ and covariance matrix $\sigma_{\beta}^2 \bm I$, where $\bm 1$ is a vector of one's and $\bm I$ is an identity matrix. Similarly, for $\bm \delta$, we assume $\bm\delta \sim N(\mu_{\delta} \bm 1, \sigma_{\delta}^2\bm I)$. For $\bm \theta$, the log spectral density at Fourier frequencies, we consider the prior from the Gaussian process, $GP(\nu(\cdot),\tau(\cdot,\cdot))$, where mean function $\nu(w)$ and covariance kernel $\tau(w_i, w_j) = (1/\tau_0^2)\text{exp}(-\rho|w_i-w_j|)$. For the hyper-parameters in the prior distribution of $\bm \theta$, we further consider $\rho \propto 1$ and $\tau_{0} \sim G(a,b)$,  where $G(a,b)$ is the Gamma distribution with mean $ab$. 
With the prior distributions, we can obtain conditional posterior distributions of the parameter given the others, which will be used to obtain the Bayesian Markov Chain Monte Carlo (MCMC) samples. Detailed derivation of the conditional posterior distributions for each parameter is relegated to the appendix.


The $k$th-step-ahead forecast based on the above model for $1 \leq k \leq M$ is obtained by estimating the conditional expectation given the data. Specifically, we use  the estimated conditional expectation given the data, i.e. $ \hat{y}_{T+k} = \hat{E}(y_{T+k}|\bm y)$ as the forecast of $y_{T+k}$ . 
With the $R$ MCMC samples from the posterior distributions, we approximate:  
$$\hat{y}_{T+k} = \frac{1}{R}\sum_{r=1}^{R}\left[X_{T+k}\hat{\bm\beta}^{(r)} + {\hat{\sigma}_{\epsilon, T+k}}^{(r)}\bm{h}^{(r)'}\tilde{\Gamma}_n^{(r)(-1)}\left(\bm y -X\hat{\bm\beta}^{(r)}\right)\right],$$
where super-script $(r)$ refers to the $r$-th  MCMC sample,  $X_{T+k}$ is the covariate vector at time $T+k$,  $\bm h = \widehat{Cov}(\bm e,e_{T+k})$ and  $\tilde{\Gamma}_n= \widehat{Cov}(\bm e)$. Note that ${\hat{\sigma}_{\epsilon, T+k}}^{(r)}$ is obtained by ${\hat{\sigma}_{\epsilon, T+k}}^{(r)}= \left(e^{\hat{\bm\eta}_{\epsilon,T+k}^{(r)}}\right)^{\frac{1}{2}}$.

\section{Empirical Results}
	
\subsection{Simulation Study}
\label{subsec:simulation}
	
A simulated data set consists of three components: covariate ($\bm X$), error ($\bm \epsilon$), and the outcome ($\bm{y}$). For simplicity, we consider two covariates, $X_1$ and $X_2$, so that $\bm X = (\textbf{1},X_1,X_2)$ has three attributes including intercept \textbf{1}. Here we assume that $X_1$ follows a bivariate mixture Gaussian distribution such that $X_1 \sim 0.5N(0,1)+0.5N(0,2)$ and that $X_2$ follows the standard exponential distribution so that $X_2 \sim Exp(1)$. For $\bm \epsilon = \bm{\sigma}_{\epsilon} \odot \bm{e}$, we construct a total of 6 different models based on two cases of volatility $\sigma_{\epsilon, t}$ and three cases of the normalized error $e_t$. 

Specifically, we let $\sigma_{\epsilon, t}$ be (1) constant over time ($\sigma_{\epsilon,t}^2\equiv 1$); (2) time-varying with a sinusoidal function, $\sigma_{\epsilon,t}^2=0.5\sin\left(\frac{\pi t}{T}\right)$. For  $e_t$, we consider (a) autoregressive model with order $2$ [AR(2)]; (b) autoregressive-moving average model with order (1,1) [ARMA(1,1)]; and (c) autoregressive conditional heteroskedasticity model with order 1 [i.e. ARCH(1)]. When $e_t \sim$ AR(2), we let the coefficients be $(a_1,a_2) = (0.5,-0.3)$, so that it satisfies ${e}_t-0.5{e}_{t-1}+0.3{e}_{t-2} = \xi{z}_t$, where ${z}_t \sim N(0,1)$, and $\xi$ is a known scaling factor $\sqrt{\frac{1-a_2}{1+a_2}\frac{var(z_t)}{(1-a_2)^2-a_1^2}} \approx 1.14$ to match its marginal variance one, i.e. $var(e_t)=1$. 
The corresponding spectral density is $\lambda(w) = 1/|1-0.5e^{i2\pi w}+0.3e^{i4\pi w}|^2$. When ${e}_t \sim$ ARMA(1,1), we set the coefficients to $(a_1,b_1)=(0.5,-0.6)$, so that ${e}_t-0.5{e}_{t-1} = -0.6z_{t-1}+{z}_{t}$ with $\lambda(w) = 1|1-0.6e^{i2\pi w}|^2/|1-0.5e^{i2\pi w}|^2$. When $e_t \sim$ ARCH(1),  we let $(\alpha_0,\alpha_1) = (0.1,0.9)$, so that it satisfies ${\sigma}_{e,t}^2 = 0.1+0.9 e_{t-1}^2$ and ${z}_t \sim N(0,1)$ with $\lambda(w) \equiv 1$. 
Additionally, we let $\bm{\beta}'=(\beta_0,\beta_1,\beta_2)=(1,2,3)$. Finally, $\bm{y}$ is generated from the model in \eqref{eqn:model2}. 

We generate each data set with the length of 205, so that the sample size for the {\it in}-{\it sample} estimation is set to $T=200$. That is, the last five sample points are used for $out$-$of$-$sample$ forecasting. We generate a total of 100 data sets for each case. For each data, the three MCMC chains with different initial values are run to generate 10,000 different sets of estimates. The results from the last 1,000 iterations are used for posterior analysis. We report the estimation and prediction results for each set of simulated data. Based on the parameter estimates, we compare our proposed model, the Bayesian Time-varying Volatility model (BTV), to some alternative approaches: 
(1) Bayesian AutorRegressive with order $1$ (BAR(1)) (2) Bayesian AutoRegressive Conditional Heteroskedasticity with order 1 (BARCH(1)) and (3) Bayesian Fixed  Volatility model (BFV) proposed by Dey {et al.} (2018) \cite{dey2018bayesian}. 


\begin{center}
	\begin{table}[h!]
		\centering
		\scalebox{0.75}{
			\begin{tabular}{|c|c|c|cc|cc|cc|}
				\hline
				\multicolumn{2}{|c|}{DGP}& \multirow{2}{*}{Fitted Model}&\multicolumn{2}{|c|}{$\hat{\beta}_0$} & \multicolumn{2}{c|}{$\hat{\beta}_1$}& \multicolumn{2}{c|}{$\hat{\beta}_2$} \\\cline{1-2}\cline{4-9}
				$\sigma_{\epsilon,t}^2$&$e_t$ & &             Bias          & RMSE           &  Bias             &    RMSE        & Bias                     &           RMSE        \\ \hline
				\multirow{12}{*}{Fixed} &\multirow{4}{*}{AR(2)}&\multicolumn{1}{|c|}{BAR(1)} &-.0656& .1219&-.0882& .2235&-.1864& .2168 \\
				&&\multicolumn{1}{|c|}{BARCH(1)} &.0306 &.1170&.0078 &.0385 &-.0141&.0826 \\
				&&\multicolumn{1}{|c|}{BFV} &-.0109 &.0925&.0022& .0277 &-.0066& .0550 \\
				&&\multicolumn{1}{|c|}{BTV} &-.0248&.0962&.0193& .0340&-.0047&.0582 \\
				\cline{2-9}
				&&\multicolumn{1}{|c|}{BAR(1)}  &-.0225 &.0998&-.0693&.0768&-.1555 &.1697\\
				&ARMA&\multicolumn{1}{|c|}{BARCH(1)}  &.0152 &.1243& .0076 &.0373&-.0055&.0851\\
				&(1,1)&\multicolumn{1}{|c|}{BFV} &-.0138& .0831&.0019 &.0267&-.0037&.0631 \\
				&&\multicolumn{1}{|c|}{BTV}  &-.0284 &.0886& .0192 &.0333&-.0031& .0656 \\			
				\cline{2-9}
				&\multirow{4}{*}{ARCH(1)}&\multicolumn{1}{|c|}{BAR(1)} &-.0122&.1168& -.0558&.0656&-.1380&.1506\\
				&&\multicolumn{1}{|c|}{BARCH(1)} &.0152& .1243  &.0076  & .0373 &-.0055& .0851 \\
				&&\multicolumn{1}{|c|}{BFV} & .0009 &.0868& .0052 &.0261&-.0130 &.0531 \\
				&&\multicolumn{1}{|c|}{BTV} &-.0190 &.0614& .0217 &.0293&-.0095& .0415\\
				\hline
				&\multirow{4}{*}{AR(2)}&\multicolumn{1}{|c|}{BAR(1)} &-.0684 &.1256& -.0661  &.0796&-.1832 &.1976 \\
				&&\multicolumn{1}{|c|}{BARCH(1)} & .0032&.1283& .0032 &.0394& .0042 &.0905 \\
				&&\multicolumn{1}{|c|}{BFV} &.0080& .1042&.0003&.0291& -.0092&.0607 \\
				&&\multicolumn{1}{|c|}{BTV} &-.0277&.0942&.0172& .0312& -.0059& .0581 \\
				\cline{2-9}
				&&\multicolumn{1}{|c|}{BAR(1)}  & -.0146 &.1004&-.0724&.0794&-.1616 &.1752\\
				Time-&ARMA&\multicolumn{1}{|c|}{BARCH(1)}  & -.0019 &.1261&-.0001  &.0403& .0035 &.0878\\
				Varying&(1,1)&\multicolumn{1}{|c|}{BFV} & -.0112 & .0918& -.0009 &.0281& -.0080 &.0677 \\
				&&\multicolumn{1}{|c|}{BTV}  & -.0312 &.0899& .0174 &.0308& -.0029&.0641 \\			
				\cline{2-9}
				&\multirow{4}{*}{ARCH(1)}&\multicolumn{1}{|c|}{BAR(1)} & -.0117  &.1155& -.0628 &.0735& -.1582 &.1752\\
				&&\multicolumn{1}{|c|}{BARCH(1)} & -.0019 & .1261  & -.0001  & .0403 & .0035 & .0878 \\
				&&\multicolumn{1}{|c|}{BFV} & .0041 &.0954& .0057 &.0303& -.0151&.0615 \\
				&&\multicolumn{1}{|c|}{BTV} & -.0161 &.0644& .0208 & .0303& -.0135 &.0461 \\
				\hline
			\end{tabular}
		}
		\caption{Bias and Root Mean Squared Error (RMSE) based on the estimates from 100 repeated sets of data for each simulation setting.
		}\label{EstimationTab}
	\end{table}
\end{center}

Table \ref{EstimationTab} shows the bias and root mean squared error (RMSE) for the estimates of $\beta_0,\beta_1$ and $\beta_2$ under the six simulation settings. The first two columns specify the structures on the volatility $\sigma^2_{\epsilon,t}$ and on the normalized stationary error $e_t$. As discussed before, we let the volatility be either fixed ($\sigma^2_{\epsilon,t}\equiv 1$) or time-varying ($\sigma^2_{\epsilon,t}=0.5\sin\left(\frac{\pi t}{T}\right)$). Secondly, the underlying structure for the stationary error $e_t$ is either AR(2), ARMA(1,1) or ARCH(1). For all six combinations, we compare the performance of our proposed approach (i.e. BTV) to those of the other three specifications (i.e. BAR(1), BARCH(1) and BFV).

Under the fixed volatility, the BAR(1) produces a relatively large bias and a large RMSE, compared to the other specifications. In contrast, the BARCH(1) generates a small bias but a relatively large RMSE than the others. In general, the BFV provides both a small bias and a small RMSE and BTV are comparable to BFV although bias for $\beta_0, \beta_1$ are larger than those of BFV. In addition, it is interesting to see that the nonparametric modeling of the spectral density (i.e. BFV, BTV) provides better outcomes than its parametric counterpart (BARCH(1)) even when the parametric specification matches the true data generating process (ARCH(1)). We have similar findings in the case of time-varying volatility as well. 

For those with time-varying volatility, BAR(1) and BARCH(1) provide relatively large RMSEs. However, BARCH(1) shows a small bias. BFV provides an overall small bias and RMSE. On the other hand, BTV shows its relative advantage in terms of RMSE for the majority of the simulation settings.

\begin{figure}[h]
	\centering
	\scalebox{1}{
		\begin{tabular}{ccc}
			\hline
			AR(2) & ARMA(1,1) & ARCH(1) \\\hline 
			\multicolumn{3}{c}{Estimated log-scaled $\sigma_{\epsilon, t}^2$} \\
			\includegraphics[width=3.5cm]{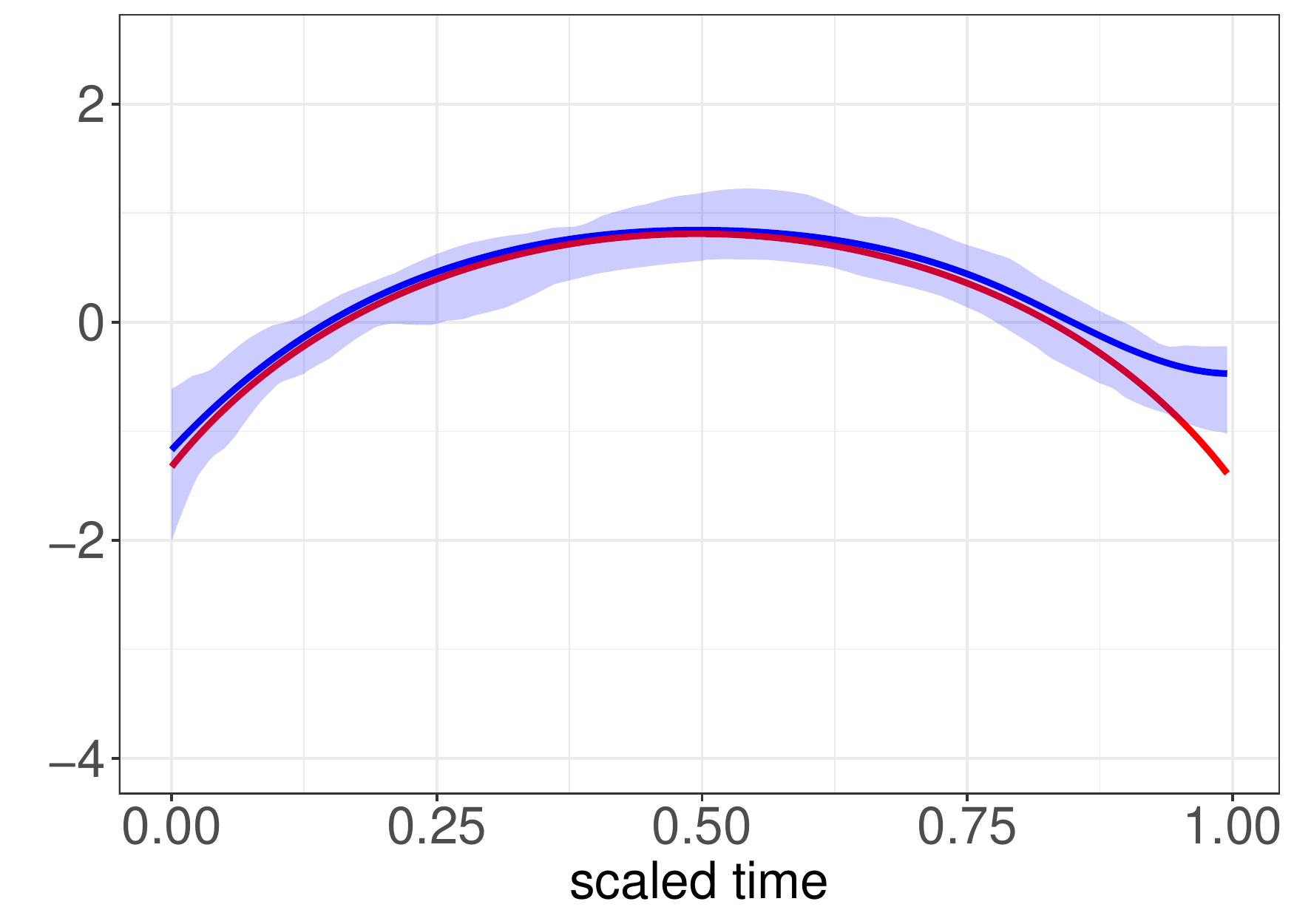} & \includegraphics[width=3.5cm]{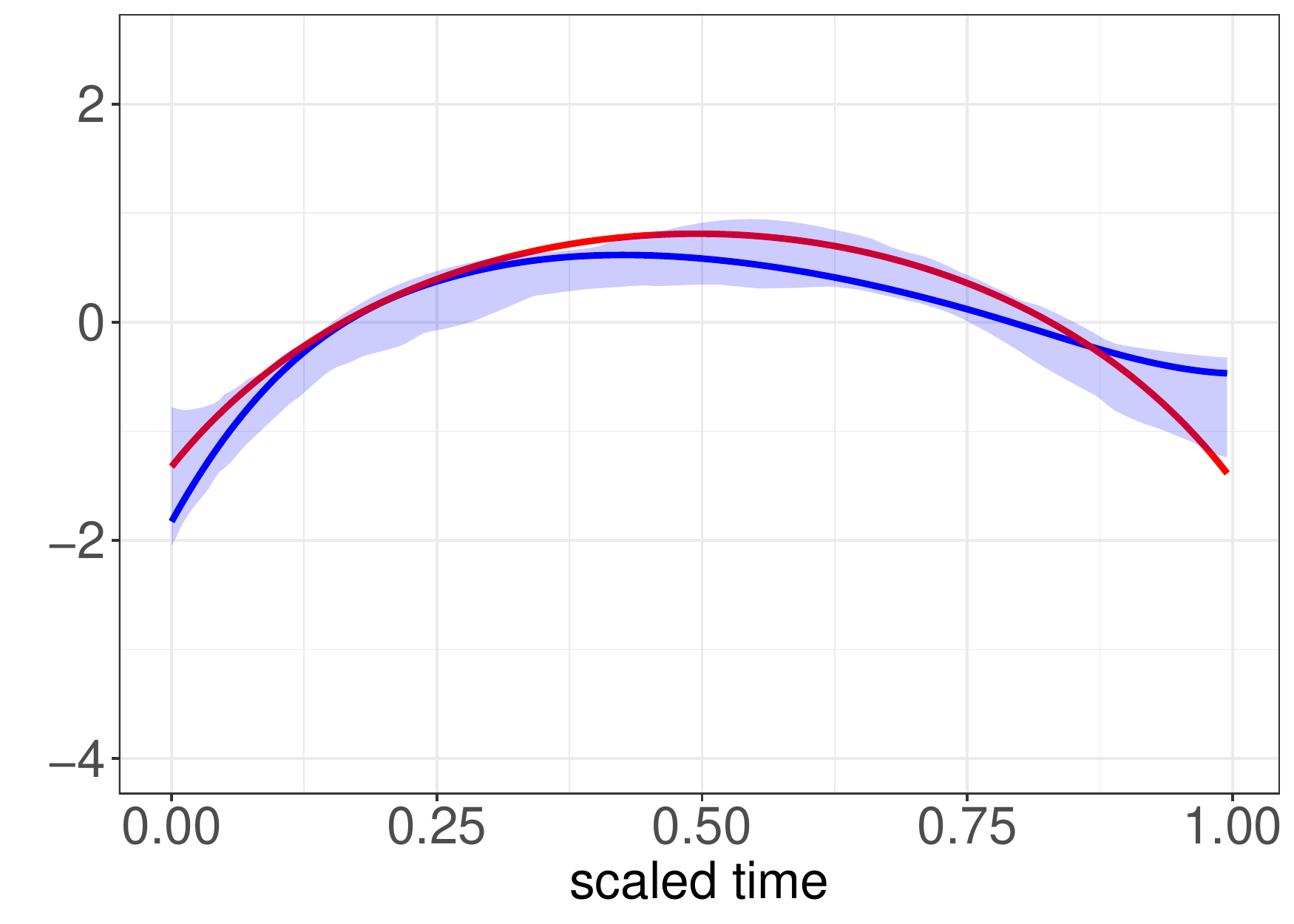} &
			\includegraphics[width=3.5cm]{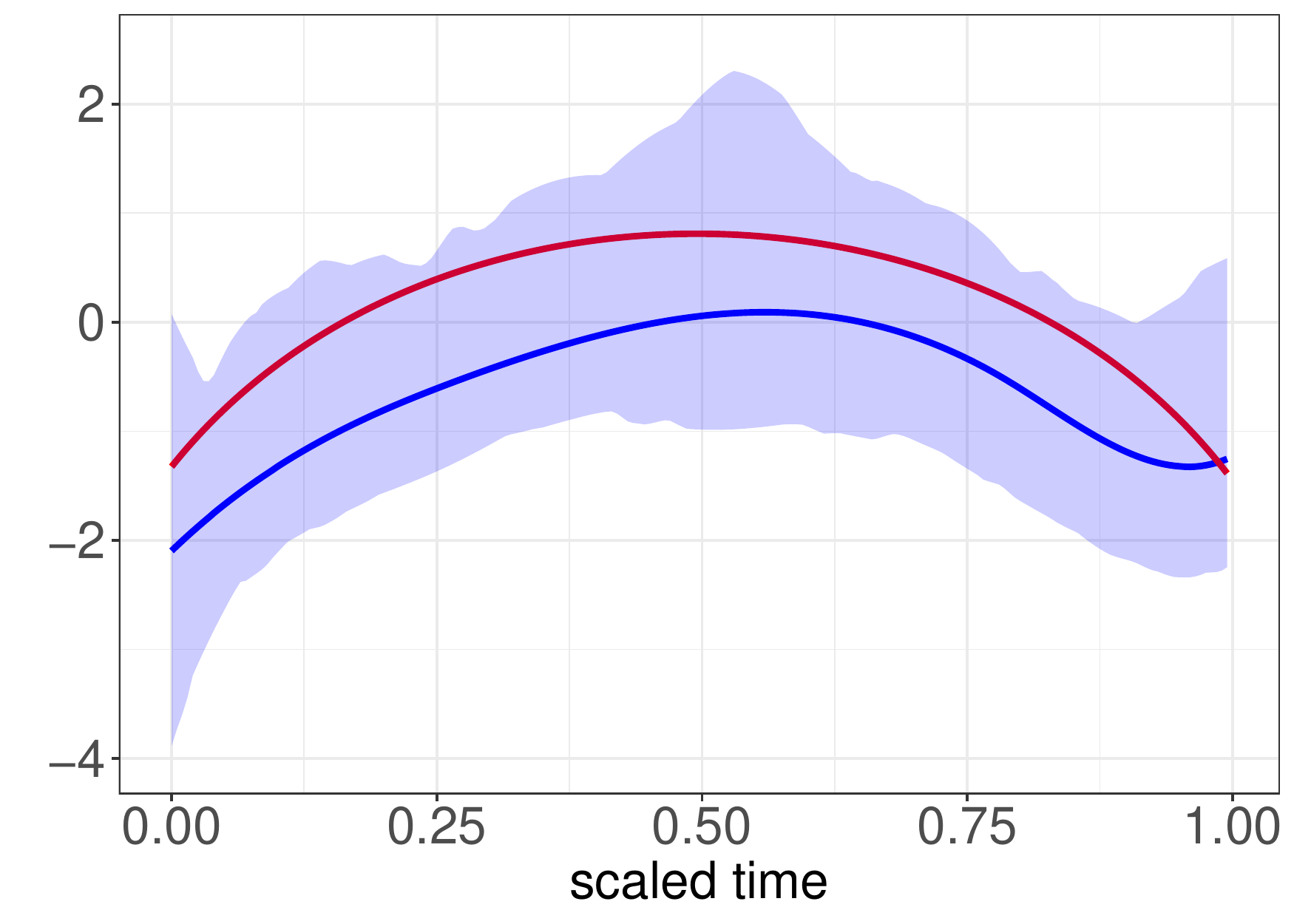}  \\ \hline
			\multicolumn{3}{c}{Estimated $\theta$}\\
			\includegraphics[width=3.5cm]{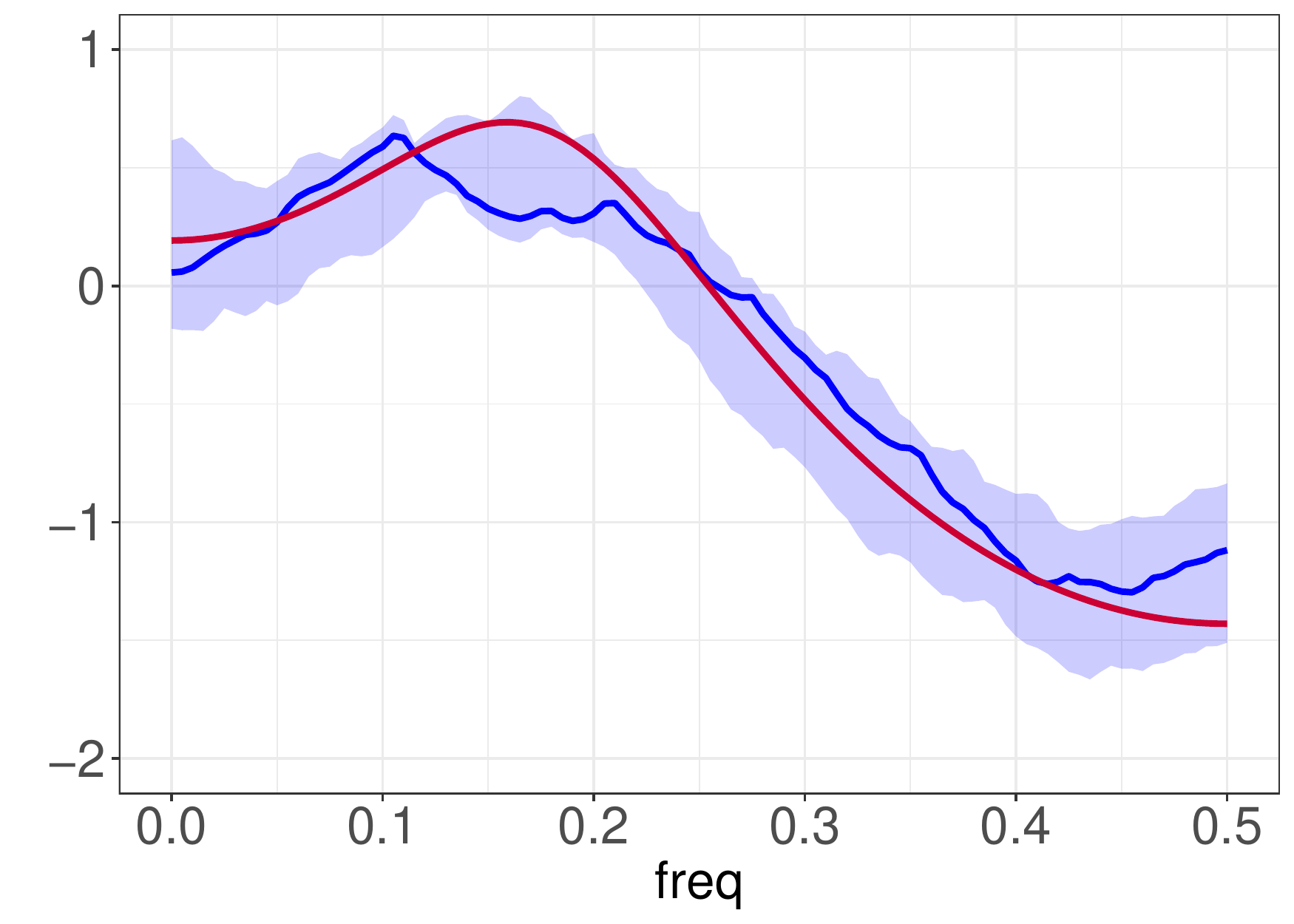} & \includegraphics[width=3.5cm]{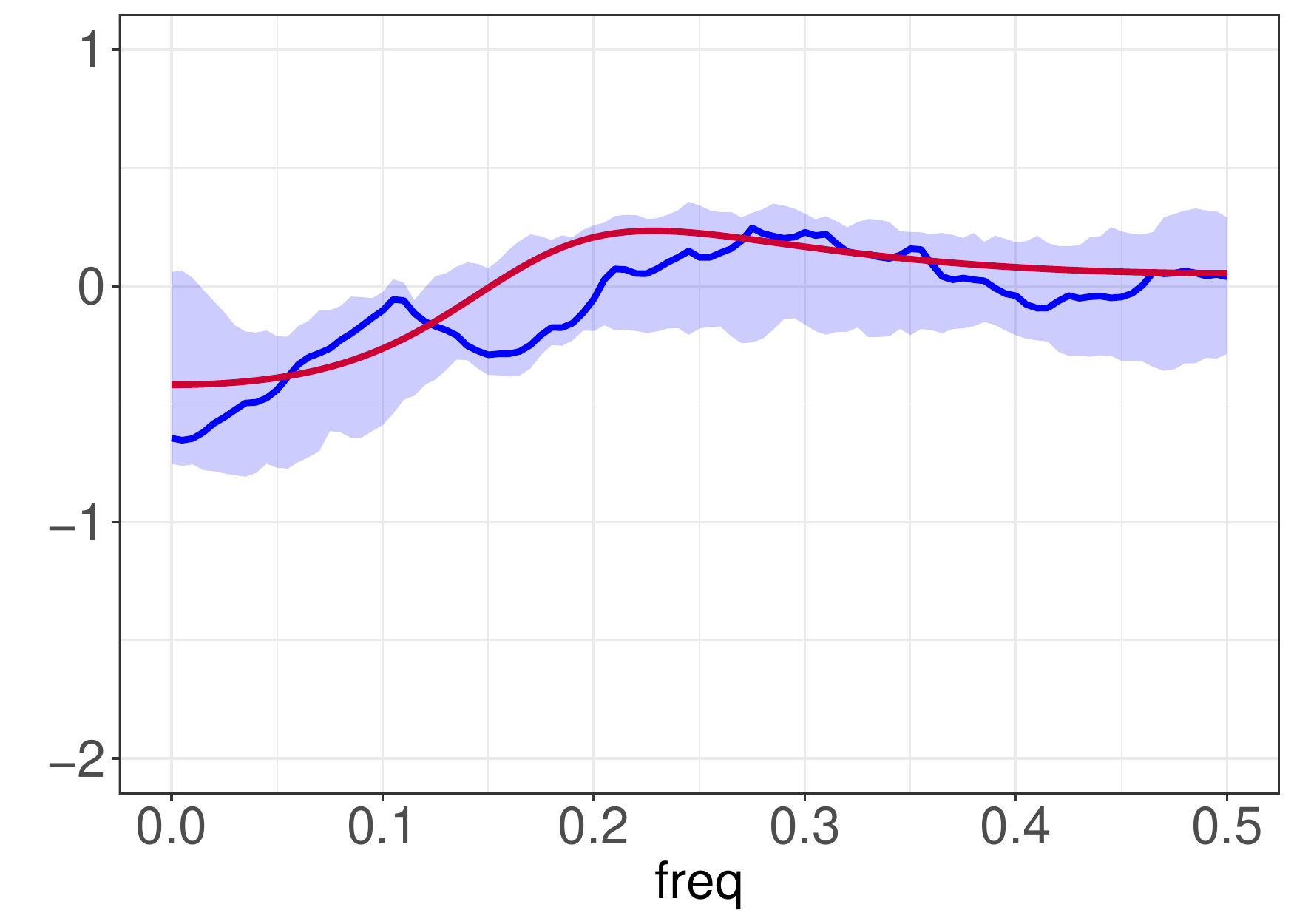} & \includegraphics[width=3.5cm]{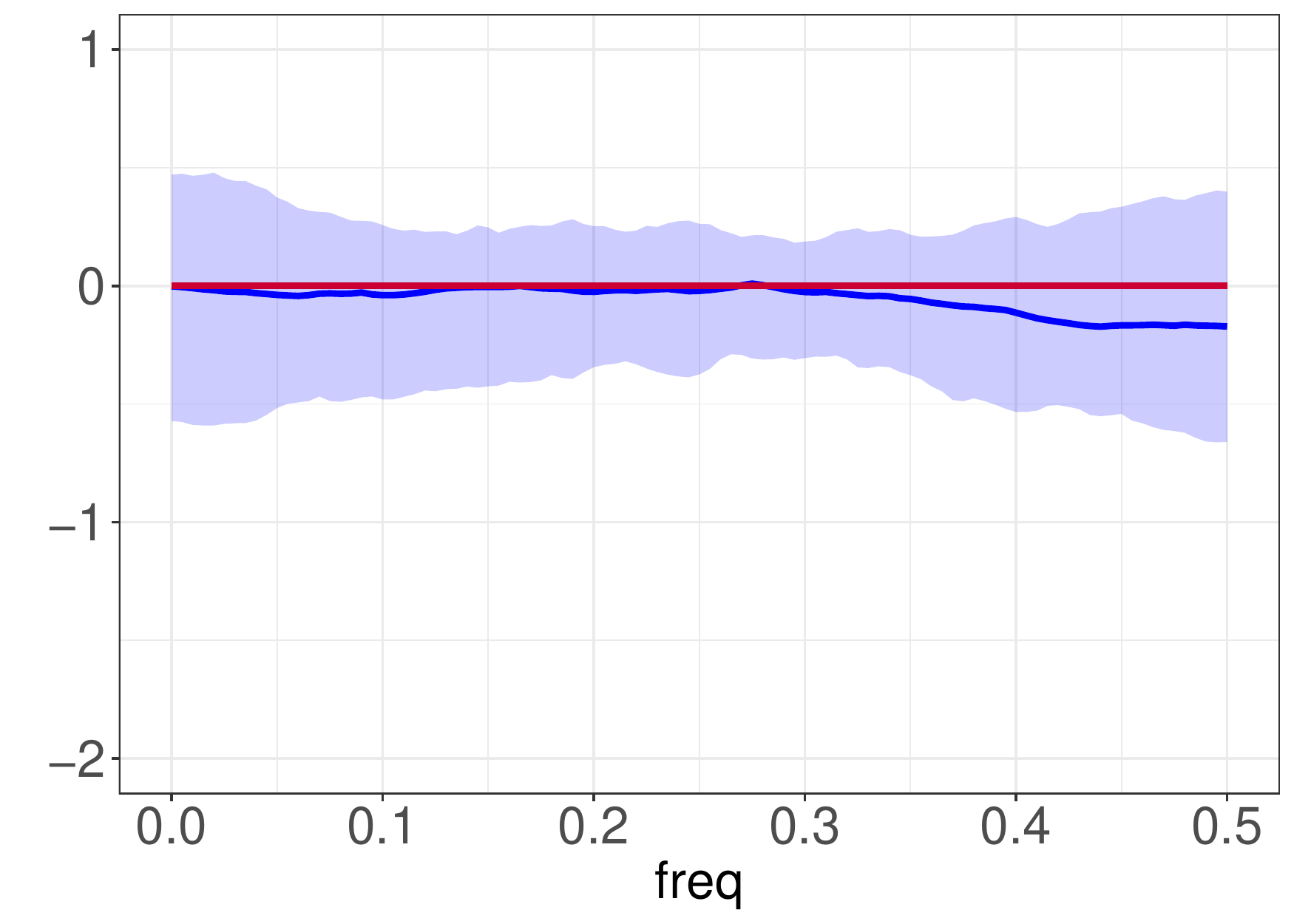}  \\\hline
		\end{tabular}
	}
	\caption{
		{
			Estimated log-scaled volatility $\sigma^2_{\epsilon,t}$ (first row) and log-spectral density $\theta$ (second row) from BTV with $95\%$ credible intervals. 
			Each column corresponds to a different data generating process.  The x-axis in the first row is the time scaled to $(0,1)$ and the x-axis in the second row  is the frequency scaled to 
			$(0,0.5)$.  The red lines indicate the true curves, blue lines indicate the estimated curves. The shaded regions indicate the 95\% credible intervals.
		}\label{EstimationFig}
	}
\end{figure}

Figure \ref{EstimationFig} shows estimated log-scaled volatility $\sigma^2_{\epsilon,t}$ and log-scaled spectral density $\theta(w)$ from the BTV when the data are simulated from the time-varying volatility ones. The estimated log-scaled $\sigma^2_{\epsilon,t}$ is somewhat biased, especially near the boundary points. However, it does capture the global pattern in the time-varying characteristics of $\sigma^2_{\epsilon,t}$ in all three cases. Regarding the spectral density estimation in the second row of Figure \ref{EstimationFig}, the density curve estimates are reasonably close to the true ones in all cases. In addition, the credible interval of the log-scaled spectral density covers the true curve of the log-scaled spectral density in all three cases.

\begin{figure}
	\begin{tabular}{ccc}
		\hline
		&Time-fixed $\sigma_{\epsilon,t}$ & Time-varying $\sigma_{\epsilon,t}$  \\\hline
		&&\\
		\multirow{-8}{*}{AR(2)}&\includegraphics[width=4.5cm]{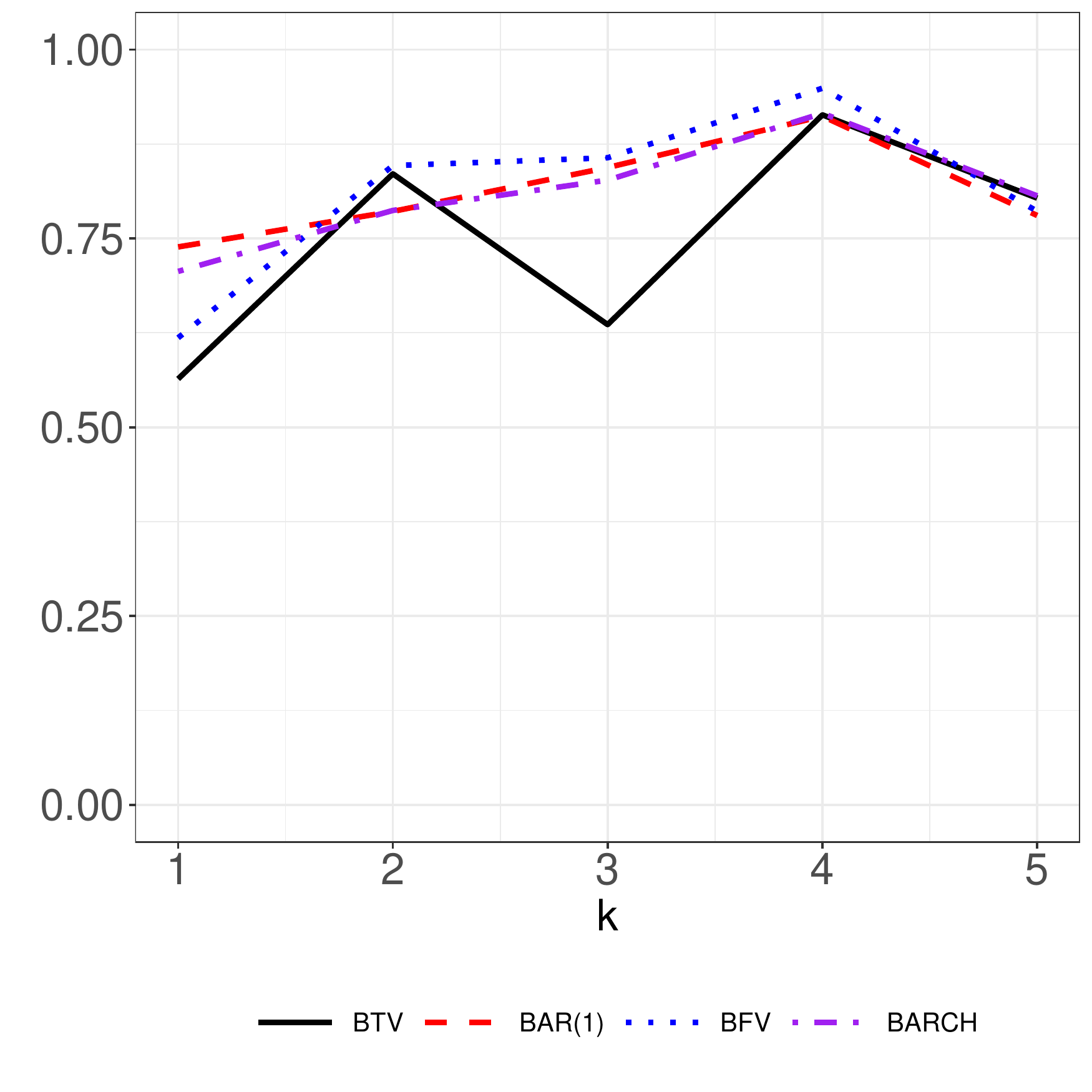} &
		\includegraphics[width=4.5cm]{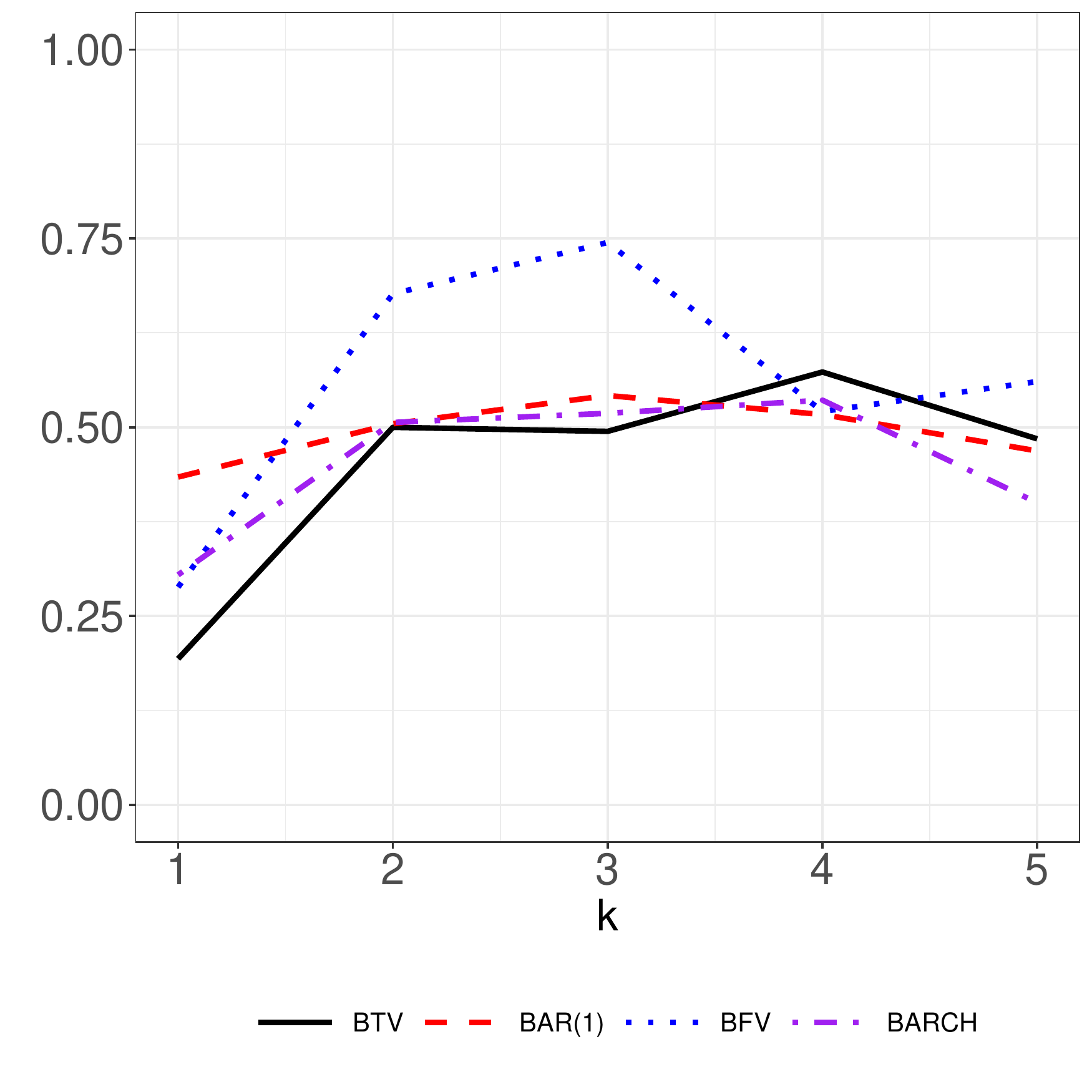}  \\ 
		\multirow{-8}{*}{\begin{tabular}[c]{@{}c@{}}ARMA\\ (1,1)\end{tabular}}&\includegraphics[width=4.5cm]{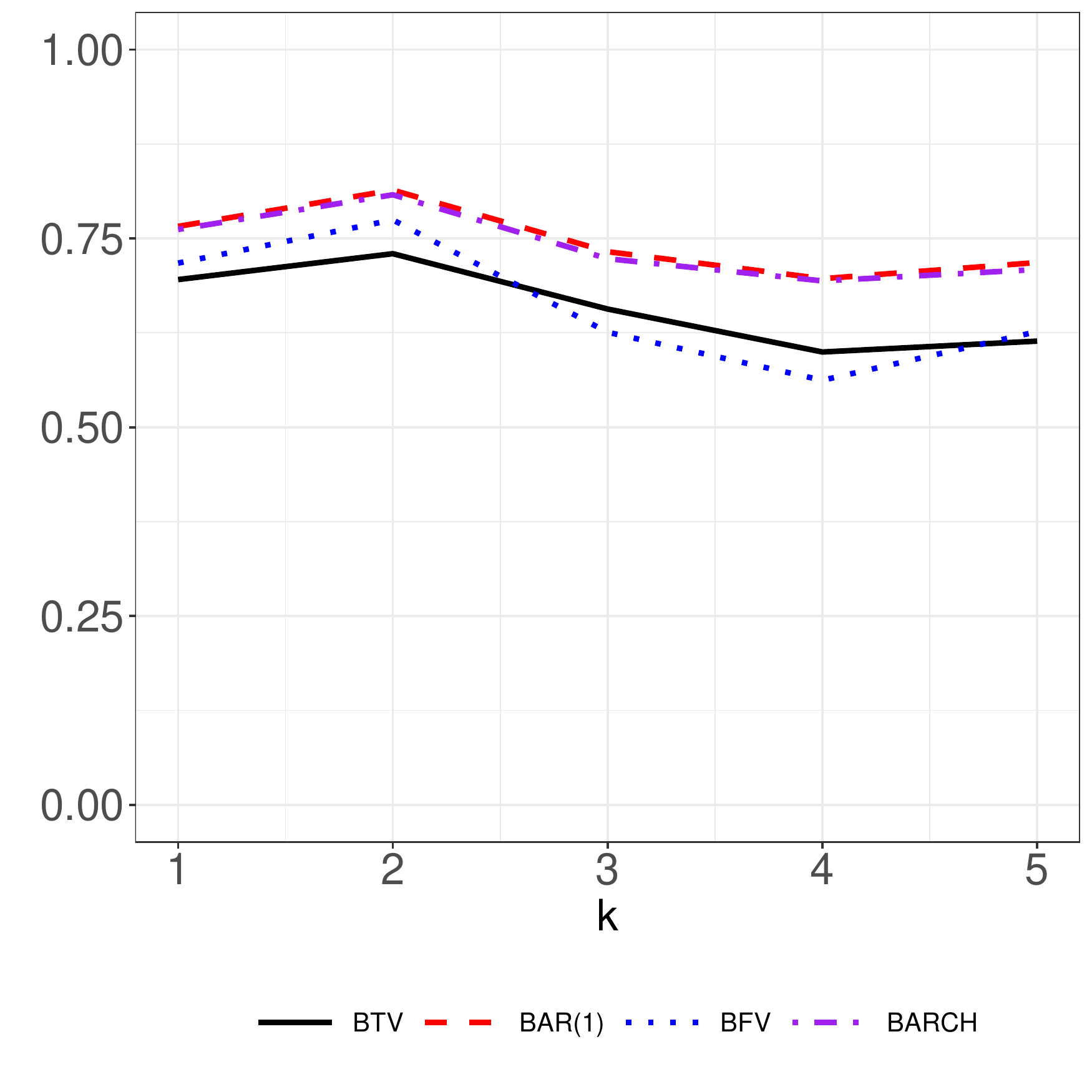} & 
		\includegraphics[width=4.5cm]{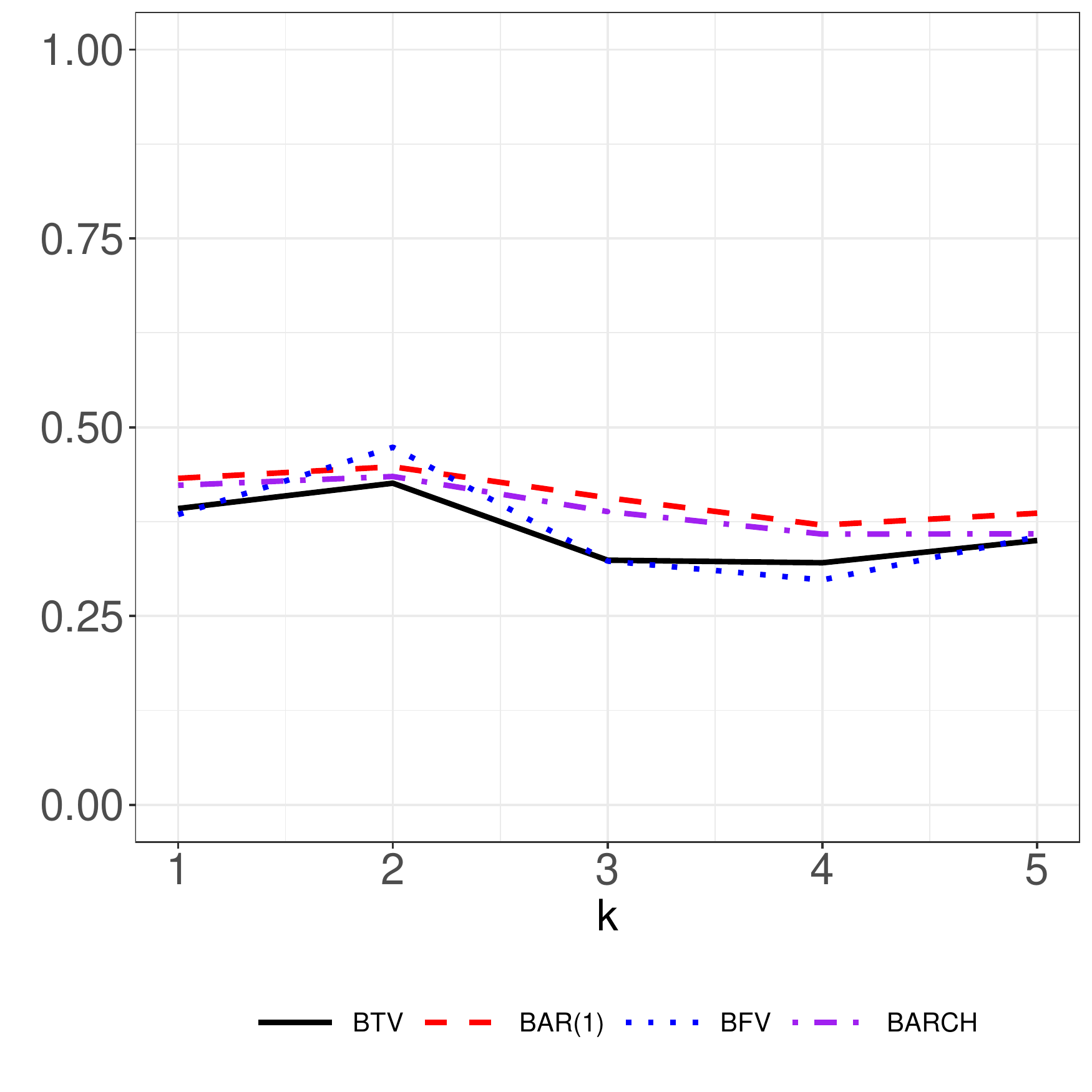}  \\
		\multirow{-8}{*}{\begin{tabular}[c]{@{}c@{}}ARCH\\ (1)\end{tabular}}&\includegraphics[width=4.5cm]{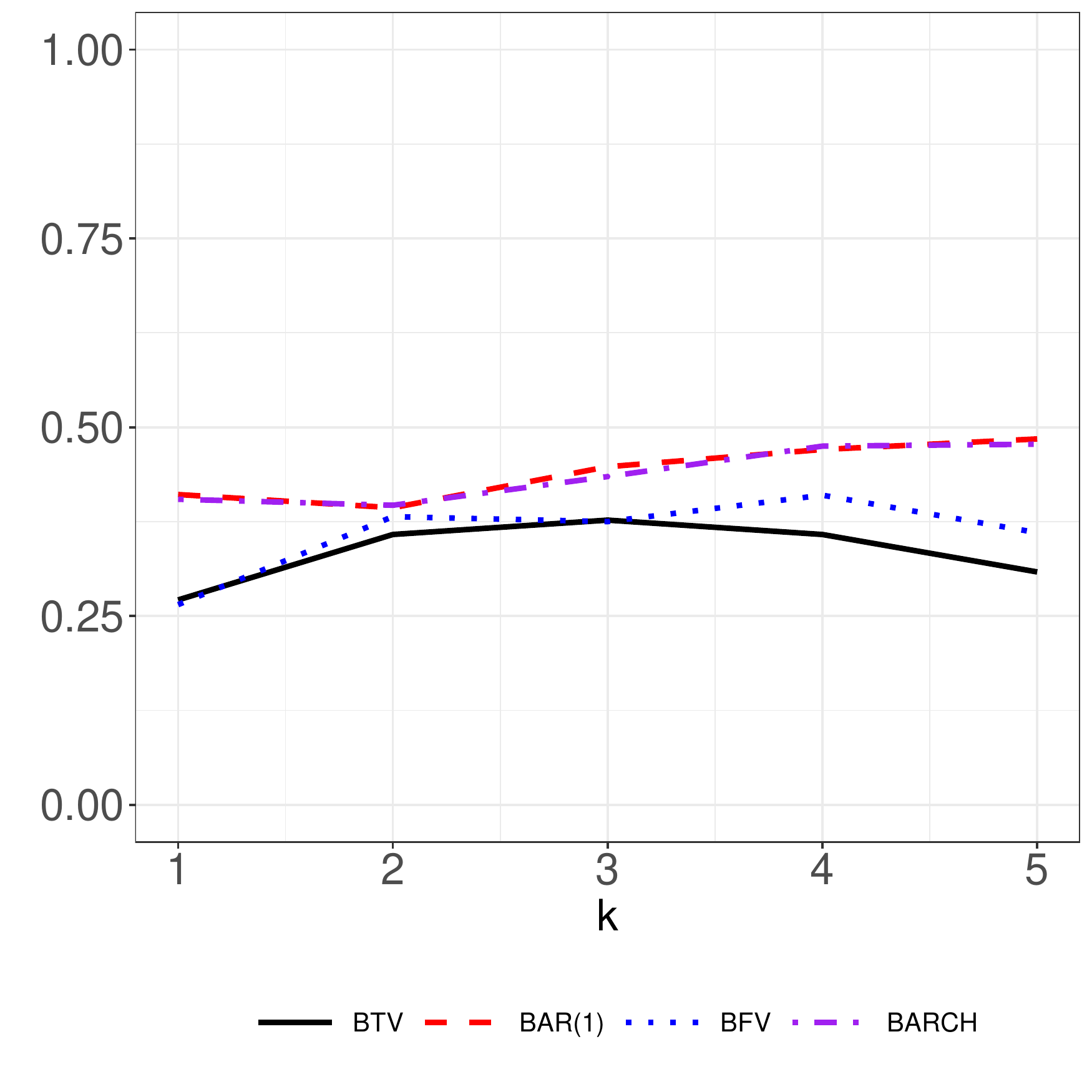} & \includegraphics[width=4.5cm]{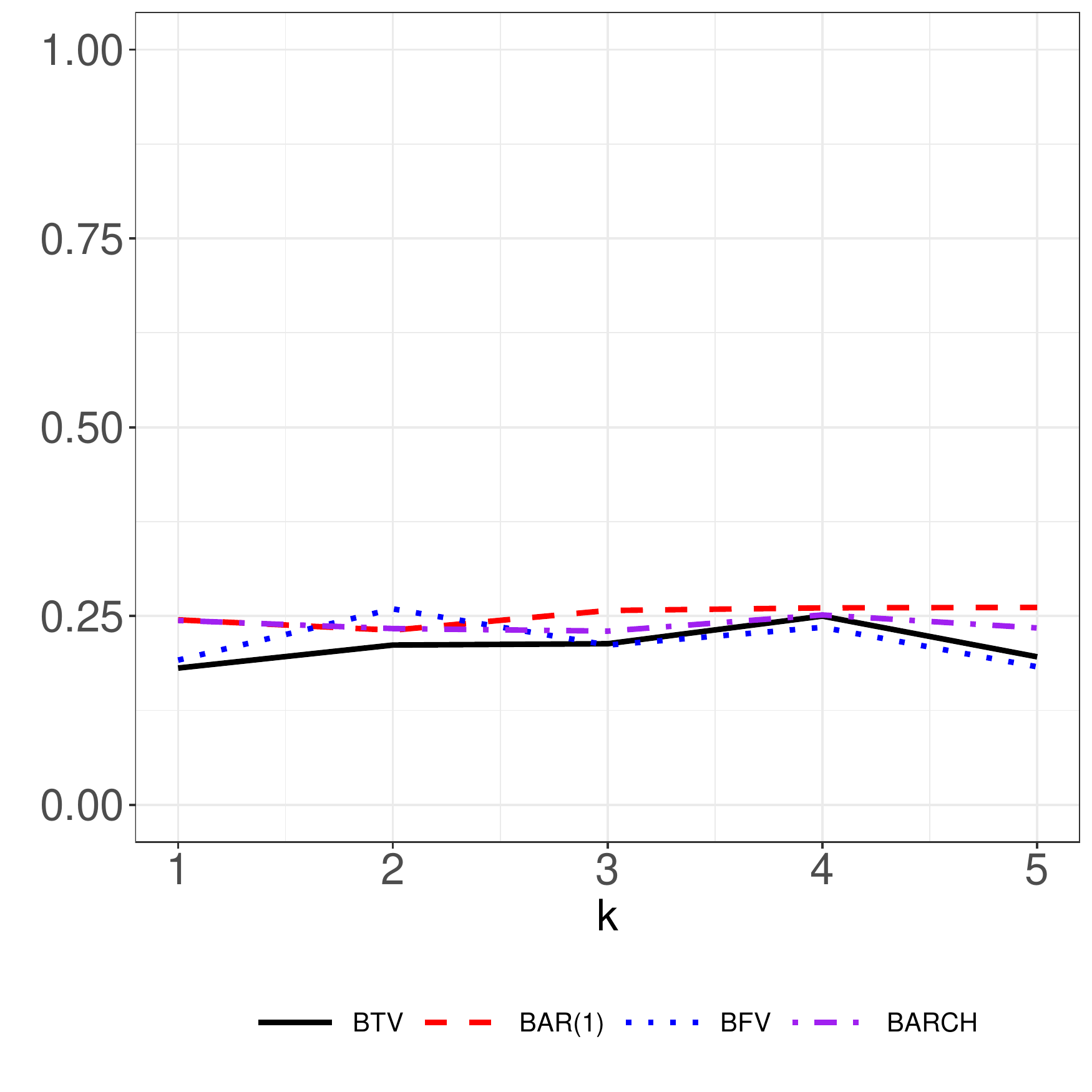}  \\ \hline
	\end{tabular}\caption{
		{
			The estimated mean squared prediction errors for $y_{n+k}$, $k=1,2,3,4,5$. The model is fitted up to time T=200 and then forecast $y_{T+k}$, $k=1,2,3,4,5$. Each row corresponds to a different data generating process. We compare the proposed model, BTV (black), to the other models: BAR(1) (red); BARCH(1) (purple); and BFV (blue). 
		}
	}\label{PredictionFig}
\end{figure}

\indent In Figure \ref{PredictionFig}, we compare the mean squared prediction errors (MSPE) of BTV to those from the other specifications at time $T+k$ for $k=1,\cdots, 5$. In each figure, the black line refers to the MSPE of BTV, the red line refers to BAR(1), the purple line refers to BARCH(1) and the blue line refers to BFV. When the true models have a time-invariant $\sigma_\epsilon$ (first column), MSPE of BTV is similar or better to those of the BFV and smaller than the other specifications.  When true models have a time-varying $\sigma_\epsilon$, BTV generally shows a lower MSPE than the rest of specifications for most of cases. 
Interestingly, the BFV and the BTV behave in a rather similar fashion, although the BFV has a higher MSPE than the BTV in the majority of the cases. 

\newpage

\subsection{Real Data Analysis}
\label{subsec:exchange rate application}

	\noindent As an application of the proposed methodology, we consider the monetary model introduced by Mark (1995) \cite{M95}: 
	
	\begin{eqnarray}
	\label{eqn:aer}
	s_{t+k} - s_{t} = \alpha + \beta(f_{t} - s_{t}) + u_{t+k}, 
	\end{eqnarray}
	where $s_t$ is log of spot exchange rate at time $t$, $s_{t+k}$ is log of spot exchange rate at time ${t+k}$. One can obtain the $k$-th-step-ahead forecast of $s_t$ from the model. Here $f_t$ is the {\it fundamental} defined by:
	$$f_t := m_t - m^*_t-\lambda(y_t - y^*_t)$$
	where $m_t$ is log of the \textit{U.S.} money supply and $y_t$ is log of the \textit{U.S.} output, while $m^*_t$ and $y^*_t$ are their foreign equivalents. For simplicity, we let $\lambda=1$ here.
		
	This study employs spot exchange rate data and their corresponding macroeconomic data, involving seven currencies of Canadian Dollar ($CAD$), Swiss Franc ($CHF$), Danish Krone ($DKK$), British Pound ($GBP$), Japanese Yen ($JPY$) and New Zealand Dollar ($NZD$) vis a vis U S Dollar ($USD$), which is the numeraire currency here. We use end-of-month observations from December 1988 through December 2016, including the 2008-2009 financial crisis period.

  We have a total of $n=336$ monthly observations for each currency. Note that $DKK$ has missing values in $m_t$ from January 2001 to March 2001. We thus impute these values by the sample average of $m_t$ in their neighborhood. We start with the first $T = 320$ observations from December 1998 until August 2015 to fit the model. That is, the estimation window length is $320$. The remaining observations are used to evaluate the model's $out$-$of$-$sample$ forecasting performance. We try four different forecast horizons, i.e. $k=1$ (monthly), $k=3$ (quarterly), $k=6$ (half yearly), and $k=12$ (yearly). Given the starting point $l$, we conduct $k$-step-ahead forecasting as follows. That is, we forecast $y_{l+T+k}$ based on the model fitted using the observations from $(l+1)$-th to $(l+T)$-th time points. 
We move $l$ from 0 to 5 for the four forecast horizons in each currency. 

	
	We compare the forecasting performance of our proposed method to those of a pack of competing models, such as the random walk without drift (RW), Bayesian Autoregressive model (BAR(1)), Bayesian Autoregressive Conditional Heteroskedasticity model (BARCH(1)) and the Bayesian fixed volatility model using the spectral density (BFV). Throughout the Bayesian $MCMC$ implementation, we run 3 chains with 10,000 iterations per chain. The last 1,000 iterations of the 3 chains are saved for estimation and inference.

\begin{table}[h]
{	\small
		\centering	
		\scalebox{1.00}{
			\begin{tabular}{|c|c|cccccc|}\hline
				Currency & $k$      &\multirow{1}{*}{RW} &\multirow{1}{*}{OLS} & \multirow{1}{*}{BAR(1)} & \multirow{1}{*}{BARCH(1)} & \multirow{1}{*}{BFV} & \multirow{1}{*}{BTV} \\ \hline
				

\multirow{4}{*}{CAD} & 1  & \textbf{0.025} & \underline{\textbf{0.023}} & \textbf{0.025}(0.40)  & \textbf{0.025}(0.54) & 0.091(0.8) & 0.088(0.00)    \\
& 3  & 0.078 & \textbf{0.049} & 0.058(0.73) & \underline{\textbf{0.036}}(0.69) & 0.063(0.9) & 0.061(1.00)    \\
& 6  & 0.106 & 0.060  & 0.059(0.20)  & 0.047(0.62) & \underline{\textbf{0.038}}(1.00)   & \textbf{0.042}(1.00)    \\
& 12 & 0.174 & \textbf{0.052} & \underline{\textbf{0.036}}(0.27) & 0.058(0.62) & 0.064(1.00)   & 0.131(1.00)    \\ \hline

\multirow{4}{*}{CHF} & 1  & 0.026 & 0.020  & 0.047(0.33) & \underline{\textbf{0.017}}(0.38) & 0.035(0.60) & \textbf{0.019}(0.25) \\
& 3  & 0.039 & 0.043 & 0.065(0.67) & \textbf{0.025}(0.54) & 0.037(1.00)   & \underline{\textbf{0.017}}(0.50)  \\
& 6  & 0.062 & 0.064 & 0.040(0.27)  & \textbf{0.025}(0.31) & 0.044(0.40) & \underline{\textbf{0.019}}(0.25) \\
& 12 & 0.035 & 0.107 & 0.034(0.53) & \underline{\textbf{0.018}}(0.77) & 0.052(0.60) & \textbf{0.023}(0.50)  \\ \hline

\multirow{4}{*}{DKK} & 1  & 0.024 & \underline{\textbf{0.018}} & 0.035(0.27) & \underline{\textbf{0.018}}(0.54) & 0.034(0.70) & \textbf{0.020}(1.00)     \\
& 3  & 0.047 & \textbf{0.031} & 0.043(0.60)  & \textbf{0.031}(0.77) & 0.041(1.00)   & \underline{\textbf{0.020}}(1.00)     \\
& 6  & 0.062 & 0.036 & 0.045(0.33) & \textbf{0.035}(0.54) & 0.043(0.50) & \underline{\textbf{0.018}}(0.25) \\
& 12 & 0.129 & 0.030  & 0.047(0.47) & \underline{\textbf{0.019}}(0.69) & 0.054(0.50) & \textbf{0.026}(0.50)  \\ \hline

\multirow{4}{*}{GBP} & 1  & 0.038 & \textbf{0.030}  & 0.051(0.33) & \underline{\textbf{0.029}}(0.23) & 0.053(0.10) & \textbf{0.030}(0.00)     \\
& 3  & \textbf{0.058} & 0.067 & 0.078(0.60)  & 0.059(0.31) & 0.062(0.20) & \underline{\textbf{0.032}}(1.00)    \\
& 6  & \textbf{0.066} & 0.120  & 0.138(0.27) & 0.102(0.31) & 0.067(0.30) & \underline{\textbf{0.034}}(1.00)    \\
& 12 & 0.145 & 0.209 & 0.207(0.53) & 0.195(0.69) & \textbf{0.075}(0.40) & \underline{\textbf{0.029}}(1.00)    \\ \hline

\multirow{4}{*}{JPY} & 1  & \textbf{0.022} & 0.029 & \underline{\textbf{0.021}}(0.53) & 0.03(0.46)  & 0.049(0.50) & 0.025(1.00)    \\
& 3  & 0.073 & 0.066 & 0.068(0.47) & 0.066(0.46) & \textbf{0.064}(0.40) & \underline{\textbf{0.032}}(1.00)    \\
& 6  & 0.114 & 0.100   & 0.103(0.47) & 0.100(0.46)   & \textbf{0.071}(0.50) & \underline{\textbf{0.034}}(0.75) \\
& 12 & 0.209 & 0.112 & 0.130(0.47)  & 0.129(0.62) & \textbf{0.070}(0.50)  & \underline{\textbf{0.030}}(0.75)  \\ \hline

\multirow{4}{*}{NZD} & 1  & 0.043 & \underline{\textbf{0.023}} & 0.041(0.80)  & \textbf{0.024}(0.62) & 0.036(0.50) & 0.036(1.00)    \\
& 3  & 0.064 & \textbf{0.034} & 0.051(0.67) & \textbf{0.034}(0.62) & 0.035(0.60) & \underline{\textbf{0.028}}(1.00)    \\
& 6  & 0.111 & 0.038 & 0.065(0.53) & 0.052(0.69) & \textbf{0.030}(1.00)    & \underline{\textbf{0.026}}(0.50)  \\
& 12 & 0.278 & 0.081 & 0.099(0.53) & 0.085(0.77) & \underline{\textbf{0.036}}(0.50) & \textbf{0.042}(1.00)    \\ \hline

			\end{tabular}
		}
		}
		\caption{Root mean square prediction error (RMSPE) of log spot exchange rate ($s_{t+k}$) for $k=1,3,6,12$-step-ahead forecasts from the proposed model (BTV) and other models (RW, OLS, BAR(1), BARCH(1), BFV) assuming the exchange rate model by Mark (1995). The ones in bold indicate the lowest RMSE for each horizon. The rest are the ratios of the model's RMSE to the lowest RMSE at the corresponding horizon. \label{log-spot-Validation-1to12month}	}
	
\end{table}

\begin{table}[h]
{	\small
		\centering	
		\scalebox{1.00}{
			\begin{tabular}{|c|c|cccccc|}\hline
				Currency & $k$      &\multirow{1}{*}{RW} &\multirow{1}{*}{OLS} & \multirow{1}{*}{BAR(1)} & \multirow{1}{*}{BARCH(1)} & \multirow{1}{*}{BFV} & \multirow{1}{*}{BTV} \\ \hline
				

\multirow{4}{*}{CAD} & 1  & 1.087 & \textbf{1.000} & 1.087          & 1.087          & 3.957          & 3.826    \\
                     & 3  & 2.167 & 1.361          & 1.611          & \textbf{1.000} & 1.750          & 1.694    \\
                     & 6  & 2.789 & 1.579          & 1.553          & 1.237          & \textbf{1.000} & 1.105    \\
                     & 12 & 4.833 & 1.444          & \textbf{1.000} & 1.611          & 1.778          & 3.639    \\ \hline

\multirow{4}{*}{CHF} & 1  & 1.529 & 1.176 & 2.765  & \textbf{1.000} & 2.059  & 1.118           \\
                     & 3  & 2.294 & 2.529 & 3.824  & 1.471          & 2.176  & \textbf{1.000}  \\
                     & 6  & 3.263 & 3.368 & 2.105  & 1.316          & 2.316  & \textbf{1.000}  \\
                     & 12 & 1.944 & 5.944 & 1.889  & \textbf{1.000} & 2.889  & 1.278           \\ \hline

\multirow{4}{*}{DKK} & 1  & 1.333 & \textbf{1.000} & 1.944 & \textbf{1.000} & 1.889  & 1.111          \\
                     & 3  & 2.350 & 1.550          & 2.150 & 1.550          & 2.050  & \textbf{1.000} \\
                     & 6  & 3.444 & 2.000          & 2.500 & 1.944          & 2.389  & \textbf{1.000} \\
                     & 12 & 6.789 & 1.579          & 2.474 & \textbf{1.000} & 2.842  & 1.368          \\ \hline

\multirow{4}{*}{GBP} & 1  & 1.310 & 1.034  & 1.759  & \textbf{1.000} & 1.828 & 1.034             \\
                     & 3  & 1.813 & 2.094  & 2.438  & 1.844          & 1.938 & \textbf{1.000}    \\
                     & 6  & 1.941 & 3.529  & 4.059  & 3.000          & 1.971 & \textbf{1.000}    \\
                     & 12 & 5.000 & 7.207  & 7.138  & 6.724          & 2.586 & \textbf{1.000}    \\ \hline

\multirow{4}{*}{JPY} & 1  & 1.048 & 1.381 & \textbf{1.000} & 1.429  & 2.333 & 1.190    \\
                     & 3  & 2.281 & 2.063 & 2.125          & 2.063  & 2.000 & \textbf{1.000}   \\
                     & 6  & 3.353 & 2.941 & 3.029          & 2.941  & 2.088 & \textbf{1.000}   \\
                     & 12 & 6.967 & 3.733 & 4.333          & 4.300  & 2.333 & \textbf{1.000}   \\ \hline

\multirow{4}{*}{NZD} & 1  & 1.870 & \textbf{1.000} & 1.783 & 1.043 & 1.565          & 1.565     \\
                     & 3  & 2.286 & 1.214          & 1.821 & 1.214 & 1.250          & \textbf{1.000}    \\
                     & 6  & 4.269 & 1.462          & 2.500 & 2.000 & 1.154          & \textbf{1.000}    \\
                     & 12 & 7.722 & 2.250          & 2.750 & 2.361 & \textbf{1.000} & 1.167             \\ \hline

			\end{tabular}
		}
		}
		\caption{ Root mean square prediction error (RMSPE) of log spot exchange rate ($s_{t+k}$) for $k=1,3,6,12$-step-ahead forecasts from the proposed model (BTV) and other models (RW, OLS, BAR(1), BARCH(1), BFV) assuming the exchange rate model by Mark (1995). The ones in bold indicate the lowest RMSPE for each horizon. The rest are the ratios of the model's RMSPE to the lowest RMSPE at the corresponding horizon. \label{ones in bold} }
	
\end{table}


 Table \ref{log-spot-Validation-1to12month} illustrates the results based on forecast error measures for the logarithm of spot return. The numbers represent the root mean squared prediction errors (RMSPE) for the forecasts, while those inside the parenthesis are the proportions that the corresponding model outperforms the random walk process in terms of the forecast error. We denote this number by RWr. That is, the closer to one the RWr is, the more frequently the corresponding model outperforms the random walk without drift.  

Empirical results in Table \ref{log-spot-Validation-1to12month} indicate that RMSPE tends to increase naturally as the forecast horizon $k$ grows. Interestingly, BTV shows relatively low RMSPE values among the competing statistical methods for the majority of currencies, namely $CHF$, $DKK$, $GBP$, $JPY$, and $NZD$. Table \ref{log-spot-Validation-1to12month} also shows that the RMSPE for the BTV is mostly either the lowest or the second lowest among all the competing ones. Except for the case of one-step-ahead forecasts, RWr of the BTV is mostly more than or equal to 50.0\% and remains relatively high in comparison to the other competing ones.


Table \ref{ones in bold} shows the {\it ratios} of each model's RMSPE to the lowest RMSPE at the corresponding horizon. Hence the best performance at each horizon is marked by {\it one in bold}. Naturally, the further away from one the ratio is, the worse the forecasting performance of the corresponding model is. As we can see from the table, the ratios for the BTV are generally close to one and frequently take on ones in bold, except for $CAD$. This means that the BTV's forecasting performance is relatively better than the other approaches/models. In particular, the BTV's ratios are typically much closer to one than those for the random walk without drift, which illustrates the relative advantage of BTV over the widely used benchmark model. By presenting the ratios instead of the original RMSPE's, one can more easily figure out the relative advantage of the BTV approach over the rest in forecasting the future spot rates. 

Our results indicate that the nonparamteric spectral-density-based method in the current study can generate more precise forecasts for foreign exchange rates because both BFV and BTV provide lower RMSPEs and higher RWrs than the alternative methods in the majority of cases. The relative advantage can be attributed to that, in contrast to the BAR(1) and BARCH(1) relying on their parametric assumptions, the BFV and BTV do not depend on any parametric covariance structure. Hence the latter could be relatively free from potential model mis-specification. Given that the BARCH(1) generally outperforms the BAR(1), modeling a time-varying volatility appears to further improve the corresponding model's forecasting performance. It should be noted, however, that it is rather hard to conclude which method is truly superior to the rest because the prediction gain for BARCH(1) over BAR(1) could be potentially due to the model's over-fitting. 

\begin{figure}[h]
	\centering
	\setlength{\tabcolsep}{-0.1em}
\begin{tabular}{cc}
		CHF
		&
		DKK \\
        \includegraphics[width=0.50\textwidth]{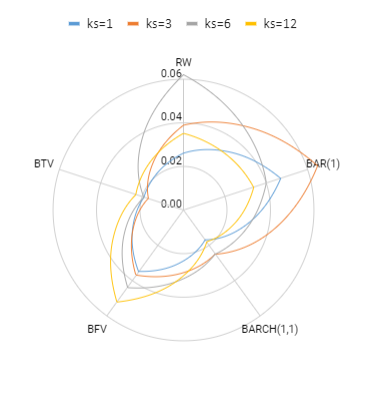}&
		\includegraphics[width=0.50\textwidth]{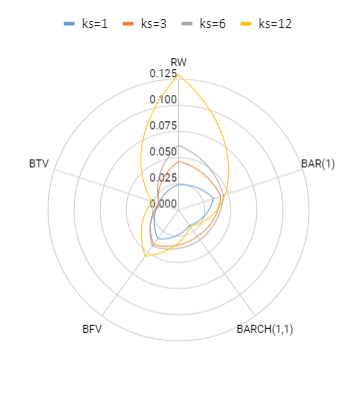} \\
		GBP
		&
		NZD \\
		\includegraphics[width=0.50\textwidth]{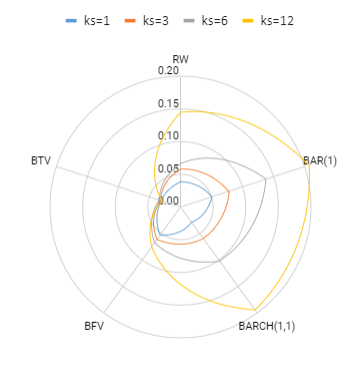}& \includegraphics[width=0.50\textwidth]{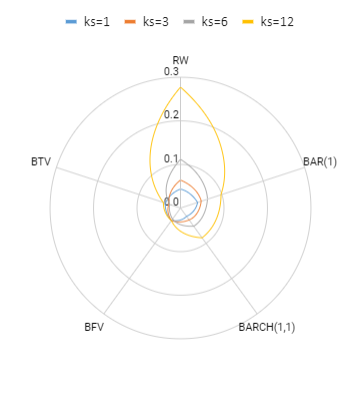}   
	\end{tabular}
	\caption{Radar charts of the root mean squared prediction errors from forecasting the log spot exchange rate for the five competing models based on four selected currencies: Switzerland Franc (CHF, top left); Danish Krone (DKK, top right); British Pound (GBP, bottom left); and New Zealand Dollar (NZD, bottom right). The different colors represent 1-month forecasting (blue), 3-month forecasting (orange), 6-month forecasting (gray), and 12-month forecasting (yellow), respectively. }\label{fig:rmse-AER}
\end{figure}

\section{Discussion}
\label{sec:discussion}

In this paper, we propose Bayesian estimation for the spectral density of an error auto-covariance in a frequency domain, without assuming any structure on the model error. The estimates are applied for the prediction of the model's dependent variable. The main motivation is to avoid any arbitrary parametric structure on the error so that estimation of the corresponding model and the forecasts of the dependent variable based on the model estimates become more precise. The approach enables us to avoid the notorious issue of nonparametric bandwidth selection by conducting the estimation in a frequency domain. The relevant Bayesian posterior distributions are derived and the estimation and prediction strategies are explained in detail. To check on its finite-sample performance, a simulation study is carefully designed and conducted. To illustrate how it works in practice, we apply the methodology to exchange rate forecasting and show that it generally outperforms the well-known benchmark in the literature, the random walk without drift. 

The current project leaves a couple of interesting issues that deserve future research on: Firstly, the current framework can be extended to accommodate {\it time-varying coefficients}. To focus on the role of time-varying volatility, our study assumes that the model coefficients are invariant over time. One can remove the restriction to make our findings more robust. Secondly, we can apply the proposed methodology to {\it density forecasting}. By forecasting the entire distribution of the dependent variable, one can forecast not only the mean but also any percentile of the underlying distribution. This would result in more comprehensive forecasts, such as density and/or interval forecasts, for the corresponding variable. Further insight can be gained by extending the current work in these and other directions.

\section*{Acknowledgement}
This work is supported by the National Research Foundation of Korea (NRF) Grant funded by the Korea government (MSIT) (No.2019R1A2C1002213,\\ No.2020R1A4A1018207).

\newpage

\bibliography{mybibfile}

\newpage

\section*{Appendix: Conditional posterior distributions}
\label{appendix}
Conditional posterior distribution of parameters for Gibbs samplers are described below.

\paragraph{Posterior of $\bm{\beta}$} ~\\
With a Gaussian prior $\bm{\beta} \sim \bm{N}(\mu_{\beta}\bm{1},\sigma_{\beta}^2\bm{I})$, we have
\begin{eqnarray*} 
	&& \bm{\beta} ~|~ \cdots ~\sim~ \bm{N}(\bm{\mu}_{*},\bm{T}_{*}),\\
	&&\bm{T}_{*} ~= ~\bm{X}^{t}\bm{\tilde{\Gamma}}^{-1}\bm{X} + \sigma_{\beta}^{-2}\bm{I}_p,\\
	&&\bm{\mu}_{*} ~=~ {\bm T}_{*}^{-1} \left( \bm{X}^{t}\bm{\tilde{\Gamma}}^{-1}\bm{Y} + \sigma_{\beta}^{-2}\mu_{\beta}\bm{I}\right),
\end{eqnarray*}

where $\bm{\tilde{\Gamma}}$ is the covariance matrix of the data constructed by $\bm{\tilde{\Gamma}} = F_{n}\Lambda_{n}F_{n}^{*}$, $F_{n}$, $\Lambda_{n}$ diagonal matrix with u-th entry $\lambda (w_{u-1})$, $F_{n}$ $n \times n $ FFT-matrix with (u,v)-th entry $q_{u.v} = \frac{1}{\sqrt{(n)}}e^{i(u-1)(v-1)2\pi/n}$, and $F_{n}^{*}$ its conjugate. This operation is performed effectively by the use of Fast Fourier Transform algorithm.

\paragraph{Posterior of $\tau_{\epsilon}$}~\\
Let $\bm{r = Y - X\beta}$. With a Gamma prior for $\tau_{\epsilon}$ we have
$$\tau_{\epsilon} ~|~ \cdots ~\sim ~ G(a_{*},b_{*}),$$
where $a_{*} = a + \frac{n_1n_2}{2}$, $b_{*} = b + \frac{1}{2}\bm{r}^t\bm{\tilde{\Gamma}}^{-1}\bm{r}$.

\paragraph{Posterior of $\bm{\theta}$}~\\
Let the normalized residual $\bm{r}_{*} = \tau_{\epsilon}^{1/2}\bm{(Y-X\beta)}$. Given $\bm r_{*}$, we  compute $\bm{\varphi}$ by (\ref{sec:2.1.2.4}). Let $\bm{\Upsilon}$ be a covariance kernel matrix of the process $\bm{\theta}$. With a Gaussian process prior of $\bm{\theta}$, we obtain
\begin{eqnarray*}
	&& \bm{\theta} ~|~ \cdots \sim \bm{N}(\bm{\nu}_{*},\bm{\Upsilon}_{*}),\\
	&& \bm{\Upsilon}_{*} ~=~ (\bm{\Upsilon}^{-1}+\bm{V}_{\psi}^{-1})^{-1},\\
	&& \bm{\nu}_{*} ~=~ \bm{\Upsilon}_{*}\bm{V}_{\psi}^{-1}(\bm{\varphi} - \bm{\kappa}_{\psi} -\bm{\nu})+\bm{\nu}
\end{eqnarray*}
where $\bm{V}_{\psi}$ = $diag\{v^2_{\psi_0},\cdots,v^2_{\psi_{n^h}}\}$, and $\bm{\kappa}_{\psi}$ = $(\kappa_{\psi_0},\cdots,\kappa_{\psi_{n^h}})^{'}$ for the assigned $n^{h}$ Fourier frequencies.

\paragraph{Posterior of $\psi$}~\\
Given the prior $P(\psi = l) = p_l$, for $l=1,\cdots,5$, 
$$P(\psi_{s}=l ~|~ \cdots) = p_l\phi_{v_l}(\varphi_{s} - \theta_{s} - \kappa_{l}),$$
for $s = 1,2,\cdots,n^{h}$, the assigned $n^{h}$ locations of Fourier frequencies.

\paragraph{Posterior of $\tau_{\theta}$}~\\
$$\tau_{\theta}~ |~ \cdots \sim G(c_{*},d_{*}),$$
where  $c_{*} = c + \frac{n^{h}}{2}$ and $d_{*} = d + \frac{1}{2}\bm{(\theta - \nu)}^t\bm{\Upsilon}^{-1}\bm{(\theta - \nu)}$.

\paragraph{Posterior of $\rho_{\theta}$} ~\\
We use a ``grid-search'' method to sample $\rho_{\theta}$. That is, for $j = 1,2$, we sample $\rho_{\theta}$ among its candidates \{$\rho_{\theta}^{(1)}$, $\rho_{\theta}^{(2)}, \cdots, \rho_{\theta}^{(M)}$\} with probability weights $\pi(\rho_{\theta}^{(l)}) \left/ \sum_{m=1}^{M}\pi(\rho_{\theta}^{(m)})\right.$, satisfying 

$$\pi(\rho_{\theta}^{(l)}) \propto \exp\left(-\frac{1}{2} log\Big|\bm{\Upsilon}^{-1}_{(\rho_{\theta}^{(l)})}\Big|\right.
\left.-\frac{1}{2}\bm{(\theta - \nu)}^t\bm{\Upsilon}^{-1}_{(\rho_{\theta}^{(l)})}\bm{(\theta - \nu)}\right),$$

\paragraph{Posterior of $\delta$}~\\
Note that $\eta_{\epsilon, t} \equiv \log(\sigma_{\epsilon, t}^2)$ is represented by cubic $B$-spline functions such that $\eta_{\epsilon, t}=\sum_{b=1}^d \delta_b \phi_b(t)$. In this model setting, let $\Phi$ be a $n \times d$ matrix of which b-th column is $\left(\phi_b(1),\cdots,\phi_b(n)\right)$. We set the i.i.d. prior on each of the coefficients $\delta_b \sim N(0,\sigma_{\delta}^2I)$, i.e. for $b=1,\cdots,d$. For $\bm \delta=(\delta_1, \cdots, \delta_d)'$, we obtain

$$\bm\delta~ |~ \cdots \propto \exp\left(-\frac{1}{2} log\Big|\Lambda^{-1}_{exp\eta}\Big|\right.\left.-\frac{1}{2}\bm{(\bm{Y} - \bm{X}\bm{\beta})}^t(\Phi^{t}\Lambda_{exp\eta}\Phi)^{-1}\bm{(\bm{Y} - \bm{X}\bm{\beta})}-\frac{1}{2\sigma_{\delta}^2}\bm{\delta}^{t}\bm{\Phi}^{t}\bm{\Phi}\bm{\delta}\right)$$

\noindent where $\Lambda_{exp\eta}$ is a diagonal matrix with u-th entry $\exp\eta(u)$. The posterior sampling is made from the Metropolis-Hastings algorithm with proposal multivariate normal of which mean is a zero vector, and standard deviation is a $d \times d$ identity matrix with all diagonal entry 1s.

\end{document}